\documentclass[aps,superscriptaddress,showpacs,floatfix,secnumarabic,prl,
 twocolumn]{revtex4}

\usepackage{amsmath,graphicx}
\usepackage{color}

\renewcommand{\vec}[1]{\boldsymbol{\mathbf{#1}}}
\newcommand{\etal}{\emph{et al}}

\begin{document}

\title{A new method of identifying self-similarity in isotropic turbulence}

\begin{abstract}
In order to analyse results for structure functions, $S_n(r)$, we
propose plotting the ratio $|S_n(r)/S_3(r)|$ against the separation $r$.
This method differs from the extended self-similarity (ESS) technique,
which plots $S_n(r)$ against $S_3(r)$, where $S_3(r) \sim r$. Using this
method in conjunction with pseudospectral evaluation of structure
functions, for the particular case of $S_2(r)$ we obtain the new result
that the exponent $\zeta_2$ decreases as the Taylor-Reynolds number
increases, with  $\zeta_2 \to 0.67 \pm 0.02$ as $R_\lambda \to \infty$. This
supports the idea of finite-viscosity corrections to the K41 prediction
for $S_2$, and is the opposite of the result obtained by ESS.
\end{abstract}

 \author{W.~D. McComb}
 \affiliation{
 SUPA, School of Physics and Astronomy,
 University of Edinburgh, Edinburgh EH9 3JZ, UK}
 \author{S.~R. Yoffe}
 \affiliation{
 SUPA, School of Physics and Astronomy,
 University of Edinburgh, Edinburgh EH9 3JZ, UK}
 \author{A. Berera}
 \affiliation{
 SUPA, School of Physics and Astronomy,
 University of Edinburgh, Edinburgh EH9 3JZ, UK}


 \maketitle

In this Letter we revisit an old, but unresolved, issue in turbulence:
the controversy that continues to surround the Kolmogorov theory (or
K41) \cite{Kolmogorov41a,Kolmogorov41b}. This controversy began  with
the publication in 1962 of Kolmogorov's `refinement of previous
hypotheses', which gave a role to the intermittency of the dissipation
rate \cite{Kolmogorov62}. From this beginning, the search for
`intermittency corrections' has grown into a veritable industry over the
years: for a general discussion, see the book \cite{Frisch95} and the
review \cite{Boffetta08}. The term `intermittency corrections' is rather
tendentious, as no relationship has ever been demonstrated between
intermittency, which is a  property of a single realization, and the
ensemble-averaged energy fluxes which underlie K41, and it is now
increasingly replaced by `anomalous exponents'. It has also been
observed by Kraichnan and others
\cite{Kraichnan74,Saffman77,Sreenivasan99,Qian00},
 that the title of K62 is misleading. It in fact represents a profoundly
different view of the underlying physics of turbulence, as compared to
K41. For this reason alone it is important to resolve this controversy.

While this search has been a dominant theme in turbulence for many
decades, at the same time there has been a small but significant number
of theoretical papers exploring the effect of finite Reynolds numbers on
the Kolmogorov exponents
\cite{Effinger87,Barenblatt98a,Qian00,Gamard00,Lundgren02}.
 All of these papers have something to say; but the last one is perhaps
the most compelling, as it appears to offer a rigorous proof of the
validity of K41 in the limit of infinite Reynolds
number. 

The controversy surrounding K41 basically amounts to: `intermittency
corrections' \emph{versus} `finite Reynolds number effects'. The former
are expected to increase with increasing Reynolds number, the latter to
decrease. In time, direct numerical simulation (DNS)
should establish the nature of high-Reynolds-number asymptotics, and so
decide between the two. In the meantime, one would like to find some way
of extracting the `signature' of this information from current simulations. 

As is well known, one way of doing this is by ESS. Our purpose here is
to propose an alternative to ESS, in which we rely on a long-established
technique in experimental physics, where the effective experimental
error can be reduced by plotting the ratio of two dependent variables:
see Chapter 3 in \cite{Bevington03}. Of course this does not work in
all cases, but only where the quantities are positively correlated. We
have verified that this is the case here and we will discuss these
secondary aspects in a more extensive paper which we intend to submit as
a regular article in due course. For the present, therefore, our
proposal is that one should plot the ratio  $|S_n(r)/S_3(r)|$ against
the separation $r$. However, here we do this only  for the case $n=2$,
since K41 \cite{Kolmogorov41a,Kolmogorov41b} involves only $S_2$ and
$S_3$, which are connected through energy conservation.

The study of turbulence structure functions (e.g. see 
\cite{vanatta70,Anselmet84}) was transformed in the mid-1990s by the 
introduction of ESS by Benzi and co-workers \cite{Benzi93,Benzi95}. Their 
method of plotting results for $S_n(r)$ against $S_3(r)$, rather than 
against the separation $r$, showed extended regions of apparent scaling 
behaviour even at low-to-moderate values of the Reynolds number, and was 
widely taken up by others e.g.  
\cite{Fukayama00,Stolovitzky93,Meneveau96,Grossmann97,Sain98}.
A key feature of this work was the implication that corrections to the 
exponents of structure functions increase with increasing Reynolds number, 
which suggests that intermittency is the dominant effect.

The longitudinal structure functions are defined as
\begin{equation}
 \label{eq:SF}
 S_n(r) = \left\langle \delta u^n_L(r) \right\rangle \ ,
\end{equation}
where the (longitudinal) velocity increment is given by
\begin{equation}
 \delta u_L(r) = \Big[ \vec{u}(\vec{x}+\vec{r},t) - \vec{u}(\vec{x},t) \Big]
 \cdot \vec{\hat{r}} \ .
\end{equation}
Integration of the K\'arm\'an-Howarth equation (KHE) 
leads, \emph{in the limit of infinite Reynolds number}, to the
Kolmogorov  `4/5' law, $S_3(r) = -(4/5)\varepsilon r$. If the
$S_n$, for $n \geq 4$, exhibit a range of power-law behaviour, then; in general, and
solely  on dimensional grounds, the structure functions of order $n$ are
expected to  take the form
\begin{equation}
 S_n(r) = C_n (\varepsilon r)^{n/3} \ .
\end{equation}
 
Measurement of the structure functions has repeatedly found a deviation
from the above dimensional prediction for the exponents. If the structure 
functions are taken to scale with exponents $\zeta_n$, thus:
\begin{equation}
 S_n(r) \sim r^{\zeta_n} \ ,
\end{equation}
then it has been found \cite{Anselmet84,Benzi95} that the difference 
$\Delta_n = \lvert n/3 - \zeta_n\rvert$ is non-zero and increases with order 
$n$.  Exponents $\zeta_n$ which differ from $n/3$ are often referred to as 
\emph{anomalous exponents} \cite{Benzi95}.

\begin{figure}[tbp]
 \begin{center}
  \includegraphics[width=0.45\textwidth]{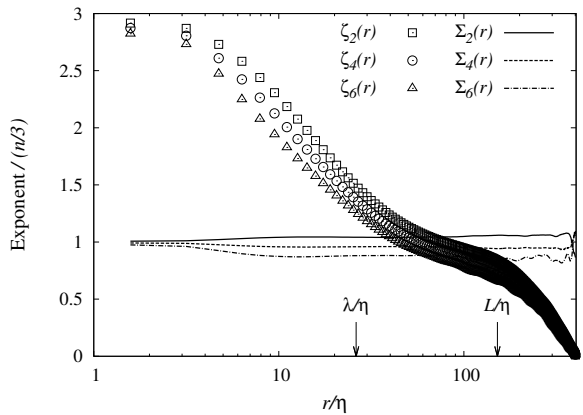}
 \end{center}
 \caption{Comparison of the local-slope exponents $\zeta_n(r)$ (symbols) for 
 $n=2, 4, 6$ with the ESS exponents $\Sigma_n(r)$ (lines). Both sets of 
 exponents were calculated from the real-space velocity field and are 
 presented here for $R_\lambda = 177$.  The separation, $r$, has been scaled 
 on the dissipation scale, $\eta = (\nu_0^3/\varepsilon)^{1/4}$.}
 \label{fig:local_grads}
\end{figure}

In order to study the behaviour of the exponents $\zeta_n$, it is usual
to make a log-log plot of $S_n$ against $r$, and measure the local slope:
\begin{equation}
 \label{eq:local_zeta}
 \zeta_n(r) = \frac{d\log{S_n(r)}}{d\log{r}} \ .
\end{equation}
Following Fukayama \etal\ \cite{Fukayama00}, the presence of a
plateau when any $\zeta_n(r)$ is plotted against $r$ indicates a constant 
exponent, and hence a scaling region.
Yet, it  is not until comparatively high Reynolds numbers are
attained that such a  plateau is found. Instead, as seen in Fig.  
\ref{fig:local_grads} (symbols),  even for the relatively large value of
Reynolds number, $R_\lambda = 177$, a scaling region cannot be
identified. (We note that Grossmann \etal\ \cite{Grossmann97} have 
argued that a
\emph{minimum} value of $R_\lambda \sim 500$ is needed for satisfactory 
direct
measurement of local scaling exponents.)

The introduction of ESS relied on the fact that $S_3$ scales with
$\zeta_3 =  1$ in the inertial range. Benzi \etal\ \cite{Benzi93}
argued that \begin{equation}
 S_n(r) \sim [S_3(r)]^{\zeta^*_n}  \ , \quad\textrm{with}\quad \zeta^*_n = 
 \zeta_n/\zeta_3 \ .
\end{equation}
$\zeta^*_n$ should then be equivalent to $\zeta_n$ in the scaling region. 

A practical difficulty led to a further step.  The statistical
convergence of odd-order structure functions is  significantly slower
than that for even-orders, due to the delicate balance  of positive and
negative values involved in the former \cite{Fukayama00}. To  overcome
this, \emph{generalized} structure functions 
have been introduced
\cite{Benzi93} (see also \cite{Stolovitzky93,Fukayama00}), 
\begin{equation}
G_n(r) = \left\langle \lvert \delta u_L(r) \rvert^n \right\rangle \sim
r^{\zeta'_n},
\end{equation}
with scaling exponents $\zeta'_n$.  The fact that $S_3 \sim r$ in the 
inertial range does not rigorously imply that $G_3 \sim r$ in the same 
range. But, by plotting $G_3(r)$ against $\lvert S_3(r)\rvert$, Benzi \etal\ 
\cite{Benzi95} showed that, for $R_\lambda = 225$ -- 800, the third-order 
exponents satisfied $\zeta'_3 \simeq 1.006\zeta_3$.  Hence, it is now 
generally assumed that $\zeta'_n$ and $\zeta_n$ are equal
(although, Fig. 2 in Belin, Tabeling and Willaime \cite{Belin96} implies 
some discrepancy at the largest length scales, and the authors note that the 
various exponents need not be the same).
Thus, by extension, $G_3$ with $\zeta'_3 = 1$, leads to
\begin{equation}
G_n(r) \sim [G_3(r)]^{\Sigma_n}, \quad\textrm{with}\quad \Sigma_n = 
\zeta'_n/\zeta'_3 \ .
\end{equation}
Benzi \etal\ \cite{Benzi93} found
that plotting their results on  this basis gave a larger scaling region.
This extended well into the dissipative lengthscales and allowed
exponents to be more easily extracted from the data. Also, Grossmann
\etal\ \cite{Grossmann97} state that the use of generalized structure 
functions is
essential to take full advantage of ESS.


There is however an alternative to the use of generalized structure
functions. This is the \emph{pseudospectral method}.
In using this for some of our work, we followed the example of
Qian \cite{Qian97,Qian99} and Tchoufag \etal\
\cite{Tchoufag12}, who obtained $S_2$ and $S_3$ from the energy and
energy transfer spectra, respectively, by means of exact quadratures.

The organization of our own work in this Letter is now as follows. We
illustrate ESS, using results from our own simulations. We also show
that our results for ESS agree closely with those of other
investigations \cite{Benzi95,Fukayama00}. We obtained these
particular results in the usual way by direct convolution sums, using a
statistical ensemble, and the generalized structure functions. We next
investigated our new method, for six Reynolds numbers spanning the range
$101.3 \leq R_\lambda \leq 335.2$. To do this we employed the
pseudospectral method \cite{Qian97,Qian99,Tchoufag12}.

We begin by illustrating the use of ESS in Fig. \ref{fig:local_grads},
where we have plotted the ESS exponents $\Sigma_n$,  for $n=2$, 4 and 6,
as lines. They may be compared to the corresponding values of $\zeta_n$,
plotted as symbols. The difference between the two sets of results is
obvious. The plots of $\zeta_n$ show no sign of scaling behaviour. In
complete contrast, the plots of $\Sigma_n$ against $r$ consist almost
entirely of plateaux, even extending well into the dissipation range,
where there would be no expectation of power-law behaviour.

It should also be noted, that we divided each exponent by the relevant 
$n/3$ in order to conveniently put the two sets of results on the  same
graph. Bearing this in mind, we see that, as $r \to 0$, the exponents 
$\zeta_n(r) \to n$ while $\Sigma_n(r) \to n/3$. This K41-type behaviour
of  the $\Sigma_n$ (which is not expected for values of $r$ in the
dissipation  range) arises because of the behaviour of the $\zeta_n$ at
small $r$, in itself a consequence of the regularity of the velocity
field. As was pointed out by Barenblatt \etal\ \cite{Barenblatt99}, who
described it as an artefact, this had been recognized from the outset by
Benzi \etal\ \cite{Benzi93}.


We used a standard pseudospectral DNS for a periodic box of side 
$L_\textrm{box} = 2\pi$, with full dealiasing performed by 
truncation according to the 2/3 rule. 
For each  Reynolds number studied, we used the same initial
spectrum ($k^4$ for the low-$k$ modes) and input rate $\varepsilon_W$.  
Stationarity was maintained using negative damping, with $\vec{f}(\vec{k},t) 
= (\varepsilon_W/2E_f) \vec{u}(\vec{k},t)$ for modes with $\lvert \vec{k} 
\rvert < k_f = 2.5$, where $E_f$ is the total energy contained in the forced 
band. The only initial condition
changed was the value assigned  to the (kinematic) viscosity, $\nu_0$.
An ensemble of $M$ realizations was generated by sampling the velocity 
every half a large-eddy turnover time, $L/U$. The simulations
are summarized in Table \ref{tbl:simulations}, and our ESS results are
plotted in Fig. \ref{fig:local_grads} for $R_\lambda = 177$, and later in Fig.
\ref{fig:exponents_summary}.

\begin{table}[tb!]
 \begin{center}
  \begin{tabular}{l|ll|llll|l}
  $R_\lambda$ & $\nu_0$ & $N$ & $\varepsilon$ & $U$ & $L/L_\text{box}$ & 
  $k_\text{max}\eta$ & $M$\\
  \hline
  42.5  & 0.01    & 128  & 0.094 & 0.581 & 0.23 & 2.34 & 101 \\
  64.2  & 0.005   & 128  & 0.099 & 0.607 & 0.21 & 1.37 & 101 \\
  101.3 & 0.002   & 256  & 0.099 & 0.607 & 0.19 & 1.41 & 101 \\
  113.3 & 0.0018  & 256  & 0.100 & 0.626 & 0.20 & 1.31 \\
  176.9 & 0.00072 & 512  & 0.102 & 0.626 & 0.19 & 1.31 & 15  \\
  203.7 & 0.0005  & 512  & 0.099 & 0.608 & 0.18 & 1.01 \\
  276.2 & 0.0003  & 1024 & 0.100 & 0.626 & 0.18 & 1.38 \\
  335.2 & 0.0002  & 1024 & 0.102 & 0.626 & 0.18 & 1.01
  \end{tabular}
 \end{center}
 \caption{A summary of the numerical simulations which have been performed. 
 The ensemble size, $M$, is given for those runs for which the ESS method 
 has been performed.}
 \label{tbl:simulations}
\end{table}

In order to examine our new proposal, we used the six runs listed in
Table \ref{tbl:simulations} with Reynolds numbers in the range $101.3
\leq R_\lambda \leq 335.2$, in conjunction with the pseudospectral
method. The second- and third-order structure functions were found from
the energy and transfer spectra, respectively, $E(k)$ and $T(k)$, using
Fourier transforms:  See Monin and Yaglom \cite{Monin75}, equations
(12.75) and (12.141$'''$). This spectral approach has the consequence
that we are now evaluating the  \emph{conventional} structure functions,
as defined by equations (1) and (2), rather than the \emph{generalized}
structure functions, $G_n(r)$, as commonly used (including by us) for
ESS.

\begin{figure}[tbp]
 \begin{center}
  \includegraphics[width=0.45\textwidth]{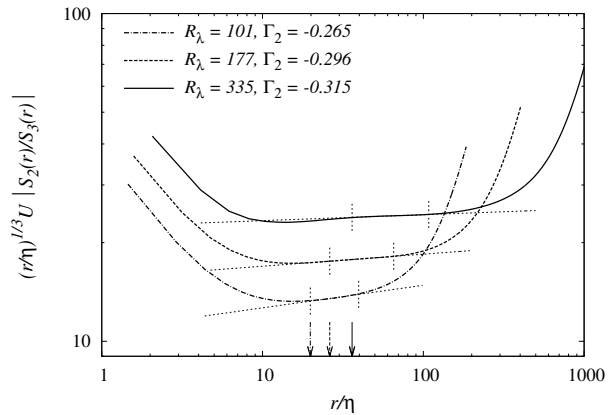}
 \end{center}
 \caption{Compensated ratio $(r/\eta)^{1/3}U\lvert 
 S_2(r)/S_3(r)\rvert$ plotted against $r$, scaled on the 
 dissipation scale, $\eta$.  K41 scaling would correspond to a 
 plateau. Arrows indicate $\lambda/\eta$, while the vertical dotted lines 
 show the region used to fit each exponent.}
 \label{fig:wdmss}
\end{figure}

\begin{figure} [tbp]
 \begin{center}
  \includegraphics[width=0.45\textwidth]{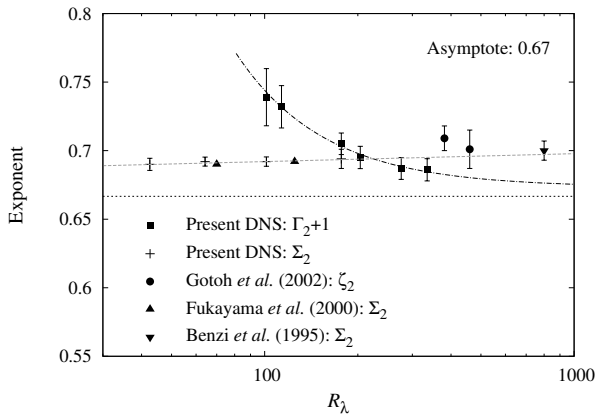}
 \end{center}
\caption{Dependence of our new exponent $\zeta_2 = \Gamma_2 
+ 1$ on Reynolds number, compared to other measured exponents: 
Fukayama \etal\ \cite{Fukayama00}, Gotoh \etal\ \cite{Gotoh02} and Benzi 
\etal\ \cite{Benzi95}. The horizontal line indicates the K41 value 
of 2/3. The dashed line indicates the fit to $\Gamma_2 + 1$, while the 
dash-dot line shows a fit to $\Sigma_2$ using our results and those of 
Fukayama \etal\ \cite{Fukayama00}.}
 \label{fig:exponents_summary}
\end{figure}

Pseudospectral calculations of the structure functions were carried
out for $R_\lambda = 101$ and 177 (and found to be comparable to the 
calculations in real space); and for the further values of the
Reynolds numbers of 113, 204, 276 and 335, which were used in Figures 
\ref{fig:wdmss} and \ref{fig:exponents_summary}.


We now arrive at the main point of this Letter, which is to
introduce a new local-scaling exponent $\Gamma_n$, which can be used
instead to determine the $\zeta_n$. We work with $S_n(r)$ and
consider the quantity $\lvert S_n(r)/S_3(r)  \rvert$.   In this
procedure, the exponent $\Gamma_n$ is defined by
\begin{equation}
 \left\lvert \frac{S_n(r)}{S_3(r)} \right\rvert \sim r^{\Gamma_n} \ ,
 \qquad\text{where}\qquad \Gamma_n = \zeta_n - \zeta_3 \ .
\end{equation}

This idea is tested,  for the case $n=2$, in Fig. \ref{fig:wdmss}.
The dimensionless quantity $U\lvert S_2(r)/S_3(r) \rvert$, where $U$ is
the rms velocity,  is plotted against $r/\eta$, for three values of 
$R_\lambda$. Note that, since K41 predicts $\Gamma_2 = -1/3$, we have
plotted a compensated form, in which we multiply the ratio by
$(r/\eta)^{1/3}$, such that K41 scaling would correspond to a plateau. From
the figure, we can see a trend towards K41 scaling as the Reynolds
number is increased.

Note that this figure also illustrates the ranges used to find
values for our new exponent $\Gamma_2$, for the following cases.
$\Gamma_2$ was fitted to the ranges $\lambda < r < c\lambda$, with $c =
2.0, 2.5$ and 3.0 for $R_\lambda = 101, 177$ and 335, respectively.

It should be emphasised that with both methods it is necessary to take
$\zeta_3=1$ in the inertial range, in order to obtain the inertial-range
value of either $\Sigma_2 = \zeta_2$ (by ESS) or $\Gamma_2 = \zeta_2 -1$
(our new method). For this reason, we plot $\Gamma_2+1$, rather than
$\Gamma_2$ in Fig. \ref{fig:exponents_summary}. From a comparison of Figs.
\ref{fig:local_grads} and \ref{fig:wdmss}, an obvious difference between
our proposed method and ESS is apparent  as $r \to 0$. This is readily
understood in terms of the regularity condition for the velocity field,
which leads to $S_n(r) \sim r^n$ as $r \to 0$
\cite{Stolovitzky93,Sirovich94}. This yields $\Gamma_n(r) + 1 \to n -
2$, whereas ESS gives  $\zeta^*_n(r) \to n/3$.

Figure \ref{fig:exponents_summary} summarizes the comparison between our
results for our new method of determining the second-order exponent and
those based on ESS (our own and others \cite{Benzi95,Fukayama00}) or on
direct measurement \cite{Gotoh02}, in terms of their overall dependence
on the Taylor-Reynolds number. In order to establish the form of the
dependence of the exponents on Reynolds number, we fitted curves to the
data points using the nonlinear-least-squares Marquardt-Levenberg
algorithm, with the error quoted being one standard error. First we
fitted a curve $\Gamma_2 + 1 = A + B R_\lambda^p$, to find the
asymptotic value $A = 0.67 \pm 0.02$. Then we fitted the curve $\Sigma_2
= C + D R_\lambda^q$, using our data and  that of Fukayama \textit{et
al} \cite{Fukayama00}. Evidently the two fitted curves show very
different trends, with results for $\Sigma_2$ increasing with increasing
Reynolds number, whereas $\zeta_2 = \Gamma_2 + 1$ decreases and
approaches 2/3 as $R_\lambda$ increases.

As we have said, the point at issue is essentially `intermittency
corrections versus finite Reynolds number effects'. The former has
received much more attention; but, in recent years, there has been a
growing interest in studying finite Reynolds number effects,
experimentally and by DNS, for the case of $S_3$: see 
\cite{Tchoufag12,Gotoh02,Antonia06} and references therein. (Although we 
note that in this case the emphasis is on the prefactor rather 
than the exponent.)

Our new result that  $\zeta_2 = \Gamma_2 +  1 \to 2/3$
may be the first indication that anomalous
values of $\zeta_2$ are due to finite Reynolds number effects.
Previously it was suggested by Barenblatt \etal\ that ESS could be
interpreted in this way \cite{Barenblatt99}, but this was disputed by
Benzi \etal\ \cite{Benzi99}.

There is much remaining to be understood about these matters and we
suggest that our new method of analysing data can help. It should, of
course, be noted that our use of $S_3$ (as evaluated by pseudospectral
methods) rather than $G_3$ (as used with ESS), may also be a factor in
our new result. In an extended account of this work (now in preparation)
we will give a full explanation of the motivation and the
circumstances under which it can be expected to work. As a matter of
interest, we conclude by noting that our analysis can also provide a
motivation for ESS and may lead to an understanding of the relationship
between the two methods. It is also the case that the pseudospectral
method could be used for the general study of higher-order structure
functions, but this awaits the derivation of the requisite Fourier
transforms.

The authors would like to thank Matthew Salewski, who read a first draft
and made many helpful suggestions. SY and AB were funded by the STFC.


\begin{thebibliography}{33}
\expandafter\ifx\csname natexlab\endcsname\relax\def\natexlab#1{#1}\fi
\expandafter\ifx\csname bibnamefont\endcsname\relax
  \def\bibnamefont#1{#1}\fi
\expandafter\ifx\csname bibfnamefont\endcsname\relax
  \def\bibfnamefont#1{#1}\fi
\expandafter\ifx\csname citenamefont\endcsname\relax
  \def\citenamefont#1{#1}\fi
\expandafter\ifx\csname url\endcsname\relax
  \def\url#1{\texttt{#1}}\fi
\expandafter\ifx\csname urlprefix\endcsname\relax\def\urlprefix{URL }\fi
\providecommand{\bibinfo}[2]{#2}
\providecommand{\eprint}[2][]{\url{#2}}

\bibitem[{\citenamefont{Kolmogorov}(1941{\natexlab{a}})}]{Kolmogorov41a}
\bibinfo{author}{\bibfnamefont{A.~N.} \bibnamefont{Kolmogorov}},
  \bibinfo{journal}{C. R. Acad. Sci. URSS} \textbf{\bibinfo{volume}{30}},
  \bibinfo{pages}{301} (\bibinfo{year}{1941}{\natexlab{a}}).

\bibitem[{\citenamefont{Kolmogorov}(1941{\natexlab{b}})}]{Kolmogorov41b}
\bibinfo{author}{\bibfnamefont{A.~N.} \bibnamefont{Kolmogorov}},
  \bibinfo{journal}{C. R. Acad. Sci. URSS} \textbf{\bibinfo{volume}{32}},
  \bibinfo{pages}{16} (\bibinfo{year}{1941}{\natexlab{b}}).

\bibitem[{\citenamefont{Kolmogorov}(1962)}]{Kolmogorov62}
\bibinfo{author}{\bibfnamefont{A.~N.} \bibnamefont{Kolmogorov}},
  \bibinfo{journal}{J. Fluid Mech.} \textbf{\bibinfo{volume}{13}},
  \bibinfo{pages}{82} (\bibinfo{year}{1962}).

\bibitem[{\citenamefont{Frisch}(1995)}]{Frisch95}
\bibinfo{author}{\bibfnamefont{U.}~\bibnamefont{Frisch}},
  \emph{\bibinfo{title}{{{T}urbulence: the legacy of {A}. {N}. {K}olmogorov}}}
  (\bibinfo{publisher}{Cambridge University Press}, \bibinfo{year}{1995}).

\bibitem[{\citenamefont{Boffetta et~al.}(2008)\citenamefont{Boffetta, Mazzino,
  and Vulpiani}}]{Boffetta08}
\bibinfo{author}{\bibfnamefont{G.}~\bibnamefont{Boffetta}},
  \bibinfo{author}{\bibfnamefont{A.}~\bibnamefont{Mazzino}}, \bibnamefont{and}
  \bibinfo{author}{\bibfnamefont{A.}~\bibnamefont{Vulpiani}},
  \bibinfo{journal}{J. Phys. A: Math. Theor.} \textbf{\bibinfo{volume}{41}},
  \bibinfo{pages}{363001} (\bibinfo{year}{2008}).

\bibitem[{\citenamefont{Kraichnan}(1974)}]{Kraichnan74}
\bibinfo{author}{\bibfnamefont{R.~H.} \bibnamefont{Kraichnan}},
  \bibinfo{journal}{J. Fluid Mech.} \textbf{\bibinfo{volume}{62}},
  \bibinfo{pages}{305} (\bibinfo{year}{1974}).

\bibitem[{\citenamefont{Saffman}(1977)}]{Saffman77}
\bibinfo{author}{\bibfnamefont{P.~G.} \bibnamefont{Saffman}}, in
  \emph{\bibinfo{booktitle}{Structure and Mechanisms of Turbulence II}}, edited
  by \bibinfo{editor}{\bibfnamefont{H.}~\bibnamefont{Fiedler}}
  (\bibinfo{publisher}{Springer-Verlag}, \bibinfo{year}{1977}),
  vol.~\bibinfo{volume}{76} of \emph{\bibinfo{series}{Lecture Notes in
  Physics}}, pp. \bibinfo{pages}{273--306}.

\bibitem[{\citenamefont{Sreenivasan}(1999)}]{Sreenivasan99}
\bibinfo{author}{\bibfnamefont{K.~R.} \bibnamefont{Sreenivasan}},
  \bibinfo{journal}{Rev. Mod. Phys.} \textbf{\bibinfo{volume}{71}},
  \bibinfo{pages}{S383} (\bibinfo{year}{1999}).

\bibitem[{\citenamefont{Qian}(2000)}]{Qian00}
\bibinfo{author}{\bibfnamefont{J.}~\bibnamefont{Qian}},
  \bibinfo{journal}{Physical Review Letters} \textbf{\bibinfo{volume}{84}},
  \bibinfo{pages}{646} (\bibinfo{year}{2000}).

\bibitem[{\citenamefont{Effinger and Grossmann}(1987)}]{Effinger87}
\bibinfo{author}{\bibfnamefont{H.}~\bibnamefont{Effinger}} \bibnamefont{and}
  \bibinfo{author}{\bibfnamefont{S.}~\bibnamefont{Grossmann}},
  \bibinfo{journal}{Z. Phys. B} \textbf{\bibinfo{volume}{66}},
  \bibinfo{pages}{289} (\bibinfo{year}{1987}).

\bibitem[{\citenamefont{Barenblatt and Chorin}(1998)}]{Barenblatt98a}
\bibinfo{author}{\bibfnamefont{G.~I.} \bibnamefont{Barenblatt}}
  \bibnamefont{and} \bibinfo{author}{\bibfnamefont{A.~J.}
  \bibnamefont{Chorin}}, \bibinfo{journal}{SIAM Rev.}
  \textbf{\bibinfo{volume}{40}}, \bibinfo{pages}{265} (\bibinfo{year}{1998}).

\bibitem[{\citenamefont{Gamard and George}(1999)}]{Gamard00}
\bibinfo{author}{\bibfnamefont{S.}~\bibnamefont{Gamard}} \bibnamefont{and}
  \bibinfo{author}{\bibfnamefont{W.~K.} \bibnamefont{George}},
  \bibinfo{journal}{Flow, turbulence and combustion}
  \textbf{\bibinfo{volume}{63}}, \bibinfo{pages}{443} (\bibinfo{year}{1999}).

\bibitem[{\citenamefont{Lundgren}(2002)}]{Lundgren02}
\bibinfo{author}{\bibfnamefont{T.~S.} \bibnamefont{Lundgren}},
  \bibinfo{journal}{Phys. Fluids} \textbf{\bibinfo{volume}{14}},
  \bibinfo{pages}{638} (\bibinfo{year}{2002}).

\bibitem[{\citenamefont{Bevington and Robinson}(2003)}]{Bevington03}
\bibinfo{author}{\bibfnamefont{P.~R.} \bibnamefont{Bevington}}
  \bibnamefont{and} \bibinfo{author}{\bibfnamefont{D.~K.}
  \bibnamefont{Robinson}}, \emph{\bibinfo{title}{{Data Reduction and Error
  Analysis for the Physical Sciences}}} (\bibinfo{publisher}{McGraw-Hill},
  \bibinfo{year}{2003}), \bibinfo{edition}{3rd} ed.

\bibitem[{\citenamefont{{van Atta} and Chen}(1970)}]{vanatta70}
\bibinfo{author}{\bibfnamefont{C.~W.} \bibnamefont{{van Atta}}}
  \bibnamefont{and} \bibinfo{author}{\bibfnamefont{W.~Y.} \bibnamefont{Chen}},
  \bibinfo{journal}{J. Fluid Mech.} \textbf{\bibinfo{volume}{44}},
  \bibinfo{pages}{145} (\bibinfo{year}{1970}).

\bibitem[{\citenamefont{Anselmet et~al.}(1984)\citenamefont{Anselmet, Gagne,
  Hopfinger, and Antonia}}]{Anselmet84}
\bibinfo{author}{\bibfnamefont{F.}~\bibnamefont{Anselmet}},
  \bibinfo{author}{\bibfnamefont{Y.}~\bibnamefont{Gagne}},
  \bibinfo{author}{\bibfnamefont{E.~J.} \bibnamefont{Hopfinger}},
  \bibnamefont{and} \bibinfo{author}{\bibfnamefont{R.~A.}
  \bibnamefont{Antonia}}, \bibinfo{journal}{J. Fluid Mech.}
  \textbf{\bibinfo{volume}{140}}, \bibinfo{pages}{63} (\bibinfo{year}{1984}).

\bibitem[{\citenamefont{Benzi et~al.}(1993)\citenamefont{Benzi, Ciliberto,
  Tripiccione, Baudet, Massaioli, and Succi}}]{Benzi93}
\bibinfo{author}{\bibfnamefont{R.}~\bibnamefont{Benzi}},
  \bibinfo{author}{\bibfnamefont{S.}~\bibnamefont{Ciliberto}},
  \bibinfo{author}{\bibfnamefont{R.}~\bibnamefont{Tripiccione}},
  \bibinfo{author}{\bibfnamefont{C.}~\bibnamefont{Baudet}},
  \bibinfo{author}{\bibfnamefont{F.}~\bibnamefont{Massaioli}},
  \bibnamefont{and} \bibinfo{author}{\bibfnamefont{S.}~\bibnamefont{Succi}},
  \bibinfo{journal}{Phys. Rev. E} \textbf{\bibinfo{volume}{48}},
  \bibinfo{pages}{R29} (\bibinfo{year}{1993}).

\bibitem[{\citenamefont{Benzi et~al.}(1995)\citenamefont{Benzi, Ciliberto,
  Baudet, and Chavarria}}]{Benzi95}
\bibinfo{author}{\bibfnamefont{R.}~\bibnamefont{Benzi}},
  \bibinfo{author}{\bibfnamefont{S.}~\bibnamefont{Ciliberto}},
  \bibinfo{author}{\bibfnamefont{C.}~\bibnamefont{Baudet}}, \bibnamefont{and}
  \bibinfo{author}{\bibfnamefont{G.~R.} \bibnamefont{Chavarria}},
  \bibinfo{journal}{Physica D: Nonlinear Phenomena}
  \textbf{\bibinfo{volume}{80}}, \bibinfo{pages}{385} (\bibinfo{year}{1995}).

\bibitem[{\citenamefont{Fukayama et~al.}(2000)\citenamefont{Fukayama, Oyamada,
  Nakano, Gotoh, and Yamamoto}}]{Fukayama00}
\bibinfo{author}{\bibfnamefont{D.}~\bibnamefont{Fukayama}},
  \bibinfo{author}{\bibfnamefont{T.}~\bibnamefont{Oyamada}},
  \bibinfo{author}{\bibfnamefont{T.}~\bibnamefont{Nakano}},
  \bibinfo{author}{\bibfnamefont{T.}~\bibnamefont{Gotoh}}, \bibnamefont{and}
  \bibinfo{author}{\bibfnamefont{K.}~\bibnamefont{Yamamoto}},
  \bibinfo{journal}{J. Phys. Soc. Japan} \textbf{\bibinfo{volume}{69}},
  \bibinfo{pages}{701} (\bibinfo{year}{2000}).

\bibitem[{\citenamefont{Stolovitzky et~al.}(1993)\citenamefont{Stolovitzky,
  Sreenivasan, and Juneja}}]{Stolovitzky93}
\bibinfo{author}{\bibfnamefont{G.}~\bibnamefont{Stolovitzky}},
  \bibinfo{author}{\bibfnamefont{K.~R.} \bibnamefont{Sreenivasan}},
  \bibnamefont{and} \bibinfo{author}{\bibfnamefont{A.}~\bibnamefont{Juneja}},
  \bibinfo{journal}{Phys. Rev. E} \textbf{\bibinfo{volume}{48}},
  \bibinfo{pages}{R3217} (\bibinfo{year}{1993}).

\bibitem[{\citenamefont{Meneveau}(1996)}]{Meneveau96}
\bibinfo{author}{\bibfnamefont{C.}~\bibnamefont{Meneveau}},
  \bibinfo{journal}{Phys. Rev. E} \textbf{\bibinfo{volume}{54}},
  \bibinfo{pages}{3657} (\bibinfo{year}{1996}).

\bibitem[{\citenamefont{Grossmann et~al.}(1997)\citenamefont{Grossmann, Lohse,
  and Reeh}}]{Grossmann97}
\bibinfo{author}{\bibfnamefont{S.}~\bibnamefont{Grossmann}},
  \bibinfo{author}{\bibfnamefont{D.}~\bibnamefont{Lohse}}, \bibnamefont{and}
  \bibinfo{author}{\bibfnamefont{A.}~\bibnamefont{Reeh}},
  \bibinfo{journal}{Phys. Rev. E} \textbf{\bibinfo{volume}{56}},
  \bibinfo{pages}{5473} (\bibinfo{year}{1997}).

\bibitem[{\citenamefont{Sain et~al.}(1998)\citenamefont{Sain, Manu, and
  Pandit}}]{Sain98}
\bibinfo{author}{\bibfnamefont{A.}~\bibnamefont{Sain}},
  \bibinfo{author}{\bibnamefont{Manu}}, \bibnamefont{and}
  \bibinfo{author}{\bibfnamefont{R.}~\bibnamefont{Pandit}},
  \bibinfo{journal}{Phys. Rev. Lett.} \textbf{\bibinfo{volume}{81}},
  \bibinfo{pages}{4377} (\bibinfo{year}{1998}).

\bibitem[{\citenamefont{Belin et~al.}(1996)\citenamefont{Belin, Tabeling, and
  Willaime}}]{Belin96}
\bibinfo{author}{\bibfnamefont{F.}~\bibnamefont{Belin}},
  \bibinfo{author}{\bibfnamefont{P.}~\bibnamefont{Tabeling}}, \bibnamefont{and}
  \bibinfo{author}{\bibfnamefont{H.}~\bibnamefont{Willaime}},
  \bibinfo{journal}{Physica D} \textbf{\bibinfo{volume}{93}},
  \bibinfo{pages}{52} (\bibinfo{year}{1996}).

\bibitem[{\citenamefont{Qian}(1997)}]{Qian97}
\bibinfo{author}{\bibfnamefont{J.}~\bibnamefont{Qian}},
  \bibinfo{journal}{Physical Review E} \textbf{\bibinfo{volume}{55}},
  \bibinfo{pages}{337} (\bibinfo{year}{1997}).

\bibitem[{\citenamefont{Qian}(1999)}]{Qian99}
\bibinfo{author}{\bibfnamefont{J.}~\bibnamefont{Qian}},
  \bibinfo{journal}{Physical Review E} \textbf{\bibinfo{volume}{60}},
  \bibinfo{pages}{3409} (\bibinfo{year}{1999}).

\bibitem[{\citenamefont{Tchoufag et~al.}(2012)\citenamefont{Tchoufag, Sagaut,
  and Cambon}}]{Tchoufag12}
\bibinfo{author}{\bibfnamefont{J.}~\bibnamefont{Tchoufag}},
  \bibinfo{author}{\bibfnamefont{P.}~\bibnamefont{Sagaut}}, \bibnamefont{and}
  \bibinfo{author}{\bibfnamefont{C.}~\bibnamefont{Cambon}},
  \bibinfo{journal}{Phys. Fluids} \textbf{\bibinfo{volume}{24}},
  \bibinfo{pages}{015107} (\bibinfo{year}{2012}).

\bibitem[{\citenamefont{Barenblatt et~al.}(1999)\citenamefont{Barenblatt,
  Chorin, and Prostokishin}}]{Barenblatt99}
\bibinfo{author}{\bibfnamefont{G.~I.} \bibnamefont{Barenblatt}},
  \bibinfo{author}{\bibfnamefont{A.~J.} \bibnamefont{Chorin}},
  \bibnamefont{and} \bibinfo{author}{\bibfnamefont{V.~M.}
  \bibnamefont{Prostokishin}}, \bibinfo{journal}{Physica D}
  \textbf{\bibinfo{volume}{127}}, \bibinfo{pages}{105} (\bibinfo{year}{1999}).

\bibitem[{\citenamefont{Monin and Yaglom}(1975)}]{Monin75}
\bibinfo{author}{\bibfnamefont{A.~S.} \bibnamefont{Monin}} \bibnamefont{and}
  \bibinfo{author}{\bibfnamefont{A.~M.} \bibnamefont{Yaglom}},
  \emph{\bibinfo{title}{{Statistical Fluid Mechanics}}}
  (\bibinfo{publisher}{MIT Press}, \bibinfo{year}{1975}).

\bibitem[{\citenamefont{Sirovich et~al.}(1994)\citenamefont{Sirovich, Smith,
  and Yakhot}}]{Sirovich94}
\bibinfo{author}{\bibfnamefont{L.}~\bibnamefont{Sirovich}},
  \bibinfo{author}{\bibfnamefont{L.}~\bibnamefont{Smith}}, \bibnamefont{and}
  \bibinfo{author}{\bibfnamefont{V.}~\bibnamefont{Yakhot}},
  \bibinfo{journal}{Phys. Rev. Lett.} \textbf{\bibinfo{volume}{72}},
  \bibinfo{pages}{344} (\bibinfo{year}{1994}).

\bibitem[{\citenamefont{Gotoh et~al.}(2002)\citenamefont{Gotoh, Fukayama, and
  Nakano}}]{Gotoh02}
\bibinfo{author}{\bibfnamefont{T.}~\bibnamefont{Gotoh}},
  \bibinfo{author}{\bibfnamefont{D.}~\bibnamefont{Fukayama}}, \bibnamefont{and}
  \bibinfo{author}{\bibfnamefont{T.}~\bibnamefont{Nakano}},
  \bibinfo{journal}{Phys. Fluids} \textbf{\bibinfo{volume}{14}},
  \bibinfo{pages}{1065} (\bibinfo{year}{2002}).

\bibitem[{\citenamefont{Antonia and Burattini}(2006)}]{Antonia06}
\bibinfo{author}{\bibfnamefont{R.~A.} \bibnamefont{Antonia}} \bibnamefont{and}
  \bibinfo{author}{\bibfnamefont{P.}~\bibnamefont{Burattini}},
  \bibinfo{journal}{J. Fluid Mech.} \textbf{\bibinfo{volume}{550}},
  \bibinfo{pages}{175} (\bibinfo{year}{2006}).

\bibitem[{\citenamefont{Benzi et~al.}(1999)\citenamefont{Benzi, Ciliberto,
  Baudet, and Ruiz-Chavarria}}]{Benzi99}
\bibinfo{author}{\bibfnamefont{R.}~\bibnamefont{Benzi}},
  \bibinfo{author}{\bibfnamefont{S.}~\bibnamefont{Ciliberto}},
  \bibinfo{author}{\bibfnamefont{C.}~\bibnamefont{Baudet}}, \bibnamefont{and}
  \bibinfo{author}{\bibfnamefont{G.}~\bibnamefont{Ruiz-Chavarria}},
  \bibinfo{journal}{Physica D} \textbf{\bibinfo{volume}{127}},
  \bibinfo{pages}{111} (\bibinfo{year}{1999}).

\end{thebibliography}
\end{document}